# Interplay between high-entropy effects and the superionic transition in fluorite oxides: A liquid-like perspective

Fengnian Zhuang[1] Gaosheng Yan[1]* Hong Chen[2] Yi Zhang[1,2] Wenshan Yu[1]* Minglong Xu[1] Shengping Shen[1]

[1]State Key Laboratory for Strength and Vibration of Mechanical Structures, Shaanxi Engineering Laboratory for Vibration Control of Aerospace Structures, School of Aerospace, Xi'an Jiaotong University, Xi'an 710049, China

[2]National Key Laboratory of Strength and Structural Integrity, Aircraft Strength Research Institute of China, Xi'an 710065, China

*corresponding to: ygs@xjtu.edu.cn (G.S. Yan) wenshan@xjtu.edu.cn (W.S. Yu)

**Abstract:** Minor actinide (MA)-bearing mixed oxide (MOX), a prominent high-entropy ceramic (HEC), is a promising candidate for advanced fast reactor fuels. Using a hybrid Monte Carlo/Molecular Dynamics (MC/MD) approach, we reveal that Np and Am segregation establishes a distinct short-range order (SRO) that induces severe lattice distortion. We uncover a competitive mechanism between localized high-entropy blocking effects and collective superionic dynamics, where the anionic kinetic and structural evolution within the 2000–2800 K range is partitioned and governed by the Tammann crossover temperature $(T_{Tam} = 2200\ \mathrm{K})$ and superionic transition temperature $(T_S = 2600\ \mathrm{K})$. In the low-temperature regime $(T < T_{Tam})$, severe lattice distortions induced by the SRO structures create localized electrostatic and strain-coupled potential wells, which spatially confine the O-ions and lead to their sluggish diffusion. At $T_{Tam} < T < T_S$, the system exhibits increased dynamical heterogeneity as thermal activation alleviates these structural constraints. Crucially, upon reaching $T_S$, these localized traps spontaneously transform into a highly interconnected, string-like <0,4,4,0>-polyhedral topological network. Driven by the maximization of configurational entropy, this structural change creates low-viscosity transport pathways that trigger confinement-enhanced diffusion. Consequently, a dramatic kinetic inversion occurs, where the O-ion mobility of SRO model surpasses that of random solid solution (RSS) model via an entropy-driven compensation mechanism. Finally, through Gibbs free energy minimization and defect analysis, structural and thermodynamic origins of the superionic transition in HEC are elucidated. This study unveils how local chemical ordering regulates ion transport under extreme conditions, providing theoretical support for the design of high-performance nuclear fuels with both superior radiation tolerance and optimized mass transfer capabilities.

## 1. Introduction

Over the decades, $(U,Pu,Th)O_2$ mixed oxide (MOX) fuels have been widely utilized in various nuclear reactors, including sodium-cooled fast reactors [1] and advanced heavy water reactors (AHWR) [2]. During nuclear fission, certain amounts of minor actinides (MAs), encompassing americium (Am) and neptunium (Np), are produced and retained within the spent fuels [1, 3]. Handling these MAs presents a significant challenge due to their high toxicity and long-lived radioactivity. To address this, MA-bearing MOX fuels, incorporating urania, plutonia, thoria and MA oxides, are fabricated and used in advanced fast reactors to achieve highly efficient MA transmutation [1, 4, 5]. Furthermore, the incorporation of MAs into MOX significantly alter its various properties [6-8], for instance, it shifts the oxygen potential and affects chemical stability, ultimately impacting overall fuel performance. Beyond terrestrial reactors, transuranic-bearing MOX fuels are also employed as radioisotope thermoelectric generators (RTGs) in space exploration [9], underscoring the versatile application prospects of MA-MOX. Crucially, by blending multiple principal cations in near-equimolar ratios, MA-bearing MOX can form stable high-entropy ceramics (HECs) governed by the entropy-stabilization rule [10]. This complex elemental composition and distribution induce unique characteristics in HECs, such as severe lattice distortion, cocktail effects, and sluggish diffusion, which collectively impart prominent thermal and mechanical stability, exceptional radiation tolerance, and superior wear resistance [11, 12]. Given these exceptional advantages, MA-bearing MOX, classified as a high-entropy oxide, warrants further investigation for its application in advanced nuclear systems. However, the strong radiotoxicity of transuranic element and the scarcity of fabricated MA-bearing MOX samples, pose significant challenges for experimental research [13]. Consequently, despite the promising potential of HEC nuclear fuels, substantial gaps remain between the practical application of high-entropy MA-MOX and our current, still infant understandings. Alternatively, molecular dynamics (MD) simulations have been successfully employed to investigate various properties of nuclear oxide compounds, including thermophysical properties [14], species diffusion behavior [15], mechanical properties [16, 17] and irradiation damages [18]. Moreover, MD simulations have proven highly effective in unveiling various key properties of HECs, such as short-range order (SRO), defect evolution and oxidation behaviors [19, 20]. During their actual in-service operation, fluorite-type MA-MOX HECs face extremely high temperatures, inevitably triggering the emergence of a distinctive phenomenon known as the superionic or Bredig transition at approximately 0.8-0.85 times the melting temperature (0.8-0.85 $T_m$) [16, 17, 21]. This transition induces pronounced anomalies in various physical properties, including species diffusivity [22], heat capacity [23], and thermal/electrical conductivity [21]. However, the fundamental origin of the superionic transition remains a subject of ongoing debate. Some studies propose that this transition arises from a timescale change driven by a collective relaxation mechanism [24], while others attribute it to the massive accumulation of point defects above the transition temperature [25, 26]. Similarly, the exact microscopic structure of superionic-state materials remains elusive. Questions persist regarding whether it manifests as a dynamic coexistence of crystalline and liquid-like sublattices [27, 28] or as a uniformly defective overall structure. From a broader perspective, the superionic state exhibits as an intermediate phase between the solid and liquid states, a phenomenon also widely observed in planetary interiors [29, 30], high-pressure ice [31], and battery electrodes [32]. Investigating the underlying mechanisms that drive the enhanced electrical conductivity, thermal conductivity, and species diffusivity in this state is crucial for designing advanced materials, including batteries, thermoelectric materials, and nuclear fuels by tailoring their mass and heat transfer capabilities. Consequently, the profound scientific and engineering value of superionic conductors has sparked growing interest in this transition across multiple research fields, spanning planetary science, condensed matter physics, nuclear fuel design, and electrochemistry. Despite the critical importance of these high-temperature phenomena, direct experimental investigations of superionic transition remain challenging due to the elevated temperatures required and necessity for precise thermal control. Additionally, the narrow temperature window in which this transition occurs makes it exceptionally difficult to capture through real-time measurements [21]. This experimental barrier is further amplified when studying HECs owing to the difficulties in sample fabrication, thereby leaving their superionic behaviors largely unexplored. Furthermore, since superionic transition is believed to be triggered by the collective long-range migration of mobile species [33], modeling this behavior requires spatial and temporal scales that far exceed the computational capability of ab initio methods. Consequently, classical MD simulations provide a reliable approach for probing these phase transition properties. Numerous studies have successfully demonstrated the efficacy of MD simulations in capturing underlying mechanisms of the superionic transition and resultant defect evolution in various materials, including actinide dioxides such as $UO_2$, $ThO_2$, and $PuO_2$ [34], actinide oxide solid solutions [16, 17, 35], and lithium compounds [36]. Considering that both high-entropy effect and the superionic transition originate from localized atomic behaviors, such as chemical short-range ordering and cooperative atomic migration, their intricate interplay is expected to significantly alter the microstructural evolution of MA-MOX HEC. Nevertheless, the complex mechanisms governing this structural and kinetic coupling lack thorough, in-depth discussion. Therefore, a systematic investigation into the complex interplay between high-entropy and superionic transition within MA-MOX HEC is highly imperative.

In this study, atomic-scale simulations incorporating a hybrid Monte Carlo/Molecular Dynamics (MC/MD) approach are employed to unravel the complex interplay between the high-entropy effects and superionic behaviors in $(Th_{0.2},U_{0.2},Pu_{0.2},Np_{0.2},Am_{0.2})O_2$ MA-MOX. Initial structural optimizations reveal a preferential segregation of Np and Am, establishing a specific short-range order (SRO) that severely distorts the cationic sublattice. This lattice distortion, quantitatively identified by atomic entropy and volumetric strain analyses, intrinsically induces sluggish diffusion within the low-temperature regime. To accurately establish the thermodynamic boundaries, the melting temperature ($T_m$) of HEC is estimated via two-phase thermodynamic (TPS) method. Building upon this, the superionic transition temperature ($T_S$) is detected by tracking the pre-melting behavior of anionic sublattice through variations in its pair distribution function (PDF). Enlightened by amorphous physics, we utilize the nonaffine squared displacement ($D^2_{\min}$) and spatial inhomogeneity parameter (*h*-factor) to decode the highly heterogeneous nature of anionic diffusion. Furthermore, through Voronoi and coordination polyhedra analysis , we elucidate the dynamic self-organization of O-ions, discovering a topological shift toward <0,4,4,0>-indexed polyhedra that assemble into continuous, string-like transport networks above $T_S$. We demonstrate that this structural evolution is thermodynamically driven by the maximization of configurational entropy and resultant minimization of Gibbs free energy. Furthermore, we identify distinct variations in cage-escape dynamics and structural relaxation, which are associated with the emergence of non-Arrhenius anionic diffusion. After that, the observed crossover in the effective liquid-like viscosity at $T_S$ further confirms this anionic kinetic inversion. In addition to $T_S$, the critical Tammann crossover temperature ( $T_{Tam}$ ) governing anionic diffusivity is precisely determined by coupling the Arrhenius and Vogel-Fulcher-Tammann (VFT) equations. Subsequently, the anionic dynamical heterogeneity is evaluated by examining the non-Gaussian parameter and characteristic timescales, revealing a dual influence from both the Tammann crossover and superionic transitions. Moreover, we investigate the generation of oxygen-Frenkel pairs (OFP), demonstrating that these critical structural evolutions are fundamentally dictated by these two distinct transition thresholds. Eventually, at the superionic transition, the system undergoes a structural avalanche where thermodynamic driving forces override localized SRO confinement, triggering massive defect generation and liquid-like diffusion.

## 2. Methods

### 2.1 Model construction and simulation setup

For the Th-U-Np-Th-Am-O HEC system, the CRG interatomic potential [37] is adopted to describe the atomic interactions among the constituent elements. Considering that the MA-MOX holds a fluorite-type structure (Fm-3m) resembling $ThO_2$, single crystalline (SC) $(Th_{0.2},U_{0.2},Pu_{0.2},Np_{0.2},Am_{0.2})O_2$ random solid solution (RSS) models are constructed by randomly substituting equal proportions of cations into the $ThO_2$ matrix using Atomsk software [38]. In these SC models, the cations occupy a face-centered cubic (FCC) sublattice, while the oxygen ions reside within the eight tetrahedral interstitial sites of the cationic sublattice. To eliminate size effects in the MD simulations, all results are averaged across three SC ($Th_{0.2}$,$U_{0.2}$,$Np_{0.2}$,$Th_{0.2}$,$Am_{0.2}$)$O_2$ models of varying dimensions: *M1* (10.84 × 10.84 × 10.84 nm³, 96,000 atoms), *M2* (16.26 × 16.26 × 16.26 nm³, 324,000 atoms) and *M3* (21.68 × 21.68 × 21.68 nm³, 768,000 atoms) (Fig. 1). All actinide metal atoms are randomly distributed within the FCC cationic sublattice. For each model, five independent random seeds are utilized to ensure statistical robustness. Subsequently, fully static and thermal relaxations are performed on these as-constructed models (detailed in Section S1 of the Supplementary Materials). Periodic boundary conditions (PBCs) are imposed in all three dimensions throughout the simulations. All MD simulations are carried out via LAMMPS software [40]. The identification of the cationic sublattice structures is achieved with the Common Neighbor Analysis (CNA) algorithm implemented in OVITO software [39].

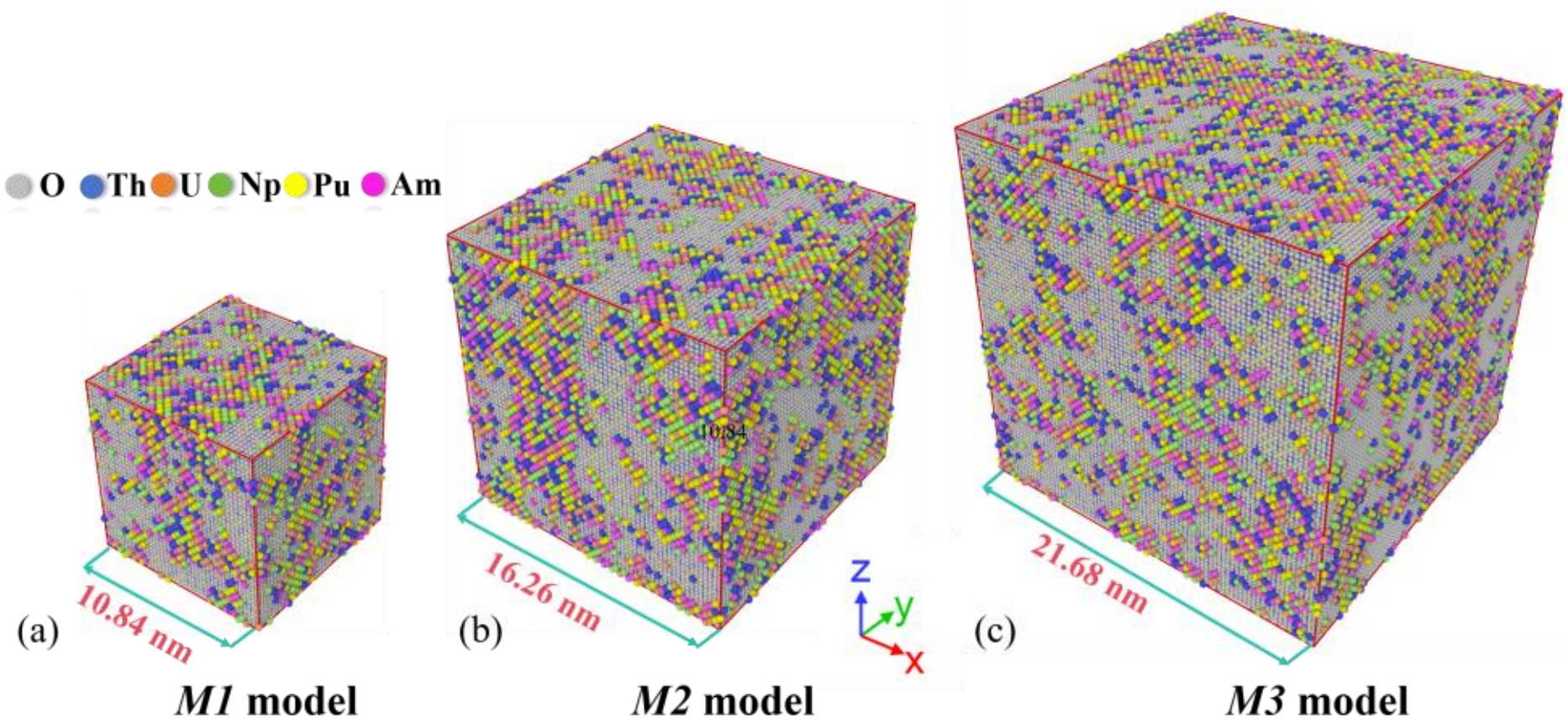

**Fig. 1** SC ($Th_{0.2}$,$U_{0.2}$,$Np_{0.2}$,$Pu_{0.2}$,$Am_{0.2}$)$O_2$ HEC models: (a) *M1* model (10.84 × 10.84 × 10.84 nm³), *M2* model (16.26 × 16.26 × 16.26 nm³), and *M3* model (21.68 × 21.68 × 21.68 nm³). The grey spheres represent O ions, while the blue, orange, green,

yellow, and pink spheres denote Th, U, Np, Pu, and Am ions, respectively.

## 2.2 Formation of the Np-Am short-range order

To determine the thermodynamically stable-state configuration of the HEC models, comprehensive hybrid MC/MD simulations are performed (detailed in Section S2 of the Supplementary Materials). The energetic and structural evolution during this hybrid MC/MD optimization is presented in Fig. 2. As exemplified by the *M*2 model in Fig. 2(a) (with other models provided in Fig. S1 of the Supplementary Materials), the potential energy undergoes a substantial decrease. Notably, Np and Am exhibit a strong tendency to segregate and cluster, ultimately forming a distinct Np-Am short-range order (SRO) structure after 1,000,000 MC steps (Fig. 2(c)), which contrasts sharply with the initial random solid solution (RSS) state (Fig. 2(b)). The Warren-Cowley (WC) parameter serves as an efficient metric for quantitatively assessing the segregation propensity of constituent elements [40]. As depicted in Fig. 2(d), the WC parameters for the Np-Np and Am-Am pairs are profoundly negative, indicating a dominant energetic drive for attraction and self-clustering. Additionally, U-U and Pu-Pu pairs exhibit moderately negative WC values, corresponding to a weaker self-clustering tendency. Conversely, the Th-Th pair yields a positive value, suggesting mild mutual repulsion. This robust Np/Am self-clustering behavior is further corroborated by their segregation propensities in ternary equimolar medium-entropy oxides (e.g., $(U_{0.33},Pu_{0.33},Am_{0.33})O_2$, $(U_{0.33},Pu_{0.33},Np_{0.33})O_2$, and $(U_{0.33},Pu_{0.33},Th_{0.33})O_2$), where Am-Am and Np-Np pairs consistently exhibit the highest degree of clustering (see Section

S3 of the Supplementary Materials). Furthermore, the local chemical ordering characterized by these WC parameters is reflected in the evolution of pair distribution functions (PDFs) for various cationic elements. As provided in Section S4 of the Supplementary Materials, the first-coordination shell peaks for Np-Np and Am-Am pairs increase significantly following the hybrid MC/MD optimization. This structural consolidation perfectly aligns with their strongly negative WC parameters. Conversely, the Th-Th peak undergoes a slight reduction, consistent with its positive WC parameter and implying mild repulsion. The PDF peak variations for all other cationic pairs similarly follow the thermodynamic trends dictated by their respective WC values. From a computational standpoint, the driving force behind this elemental segregation originates primarily from variations in the many-body perturbation terms fitted for different cations within the CRG potential, particularly regarding the local charge density and embedding energy functions [37]. Physically, these parameter disparities capture the intrinsic physicochemical differences among the actinides, including varying degrees of 5*f* electron localization, ionic radii, and electronegativity differences [41-43] (detailed in Section S5 of the Supplementary Materials).

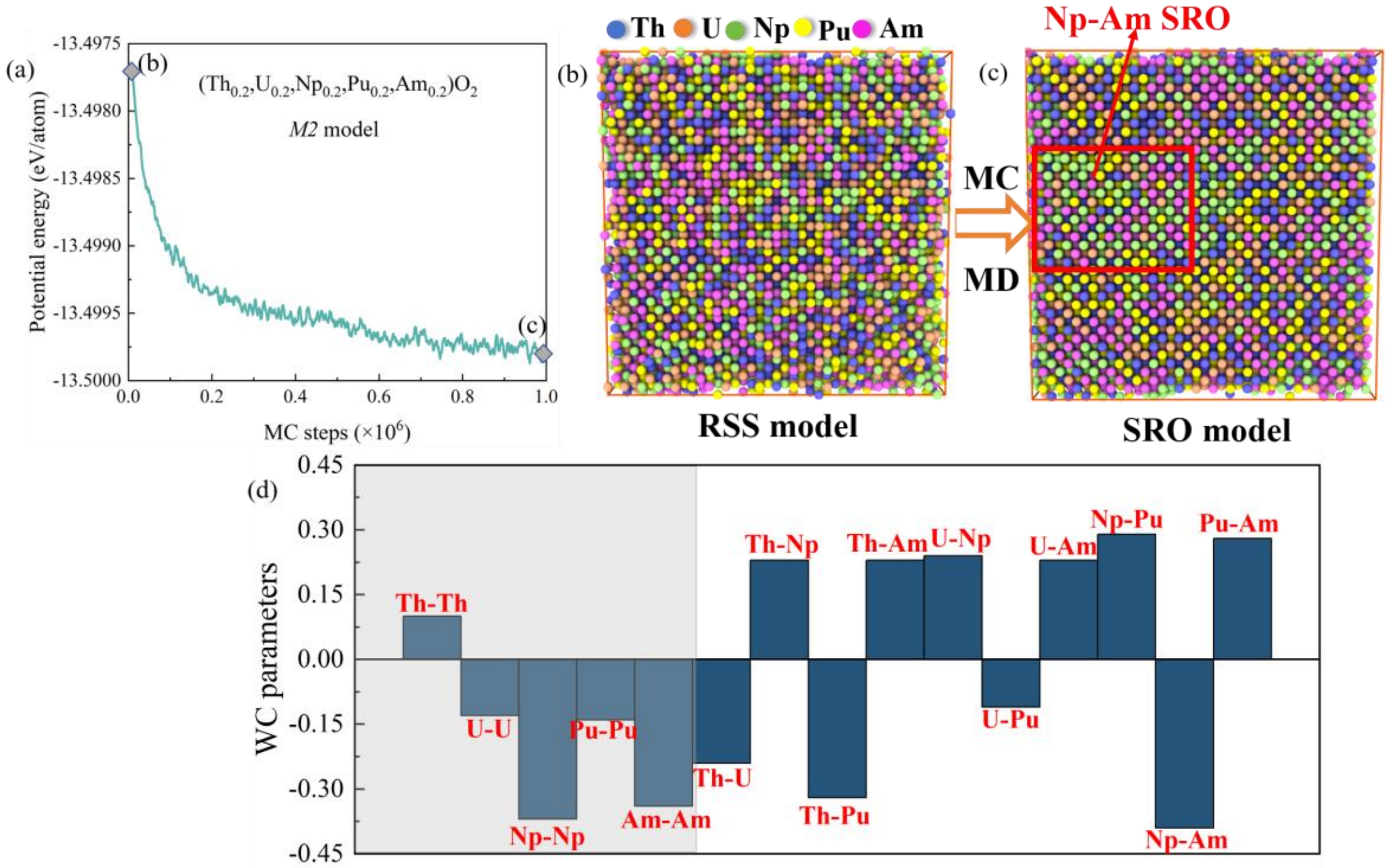

**Fig. 2** (a) Variations of potential energy per atom versus MC steps during the hybrid MC/MD optimization for SC *M2* HEC model. Atomic snapshots of the (b) initial RSS and (c) optimized SRO configurations. (d) Warren-Cowley (WC) parameters for various actinide element pairs, indicating distinct clustering and exclusion tendencies within the cationic sublattice.

## 2.3 Determination of the melting temperature

Prior to characterizing the high-temperature physical properties, an accurate estimation of the melting temperature ($T_m$) is a crucial prerequisite. The two-phase simulation (TPS) method, established as a robust approach for precisely calculating $T_m$ in actinide oxides [2, 44], is employed to determine the melting point of $(Th_{0.2},U_{0.2},Pu_{0.2},Np_{0.2},Am_{0.2})O_2$ HEC. The TPS configuration comprises a 6 × 6 × 12 supercell, with half initially set to a solid phase at 1500 K and the other half to a liquid phase at 5000 K (Fig. 3(c) and (d)-I, III). During the TPS, the simulation box is heated from 1500 K to 4000 K in 25 K intervals under the NPT ensemble, with a 1 ns equilibration at each temperature (see Section S6 of the Supplementary Materials for details). The variations in density and enthalpy increment ( $H(T)-H(300\text{ K})$ ) are presented in Fig. 3 to accurately determine the $T_m$. The system density exhibits a monotonic decrease across the entire temperature range, reflecting continuous thermal expansion. A sharp discontinuity in density is observed between 3225 K and 3275 K (Fig. 3(a)), suggesting the onset of melting. Similarly, the enthalpy increment shows a distinct jump near 3250 K (Fig. 3(b)). These trends are consistent with reported melting behavior in $UO_2$ [45] and MOX [2], serving as reliable benchmarks for $T_m$ identification. As shown in Fig. 3(c) and (d), the simulation box reverts to a pure solid phase at $T=3225$ K, whereas it fully transitions to a liquid state at $T=3250$ K after 1 ns of equilibration. The migration of the solid-liquid interface is further verified via Common Neighbor Analysis (CNA) of the cationic sublattice (Fig. 3(c) and (d) III, IV). By corroborating these quantitative thermodynamic curves with the atomic-scale snapshots, the melting temperature $T_m$ of the $(Th_{0.2},U_{0.2},Pu_{0.2},Np_{0.2},Am_{0.2})O_2$ HEC is determined to be approximately 3250 K.

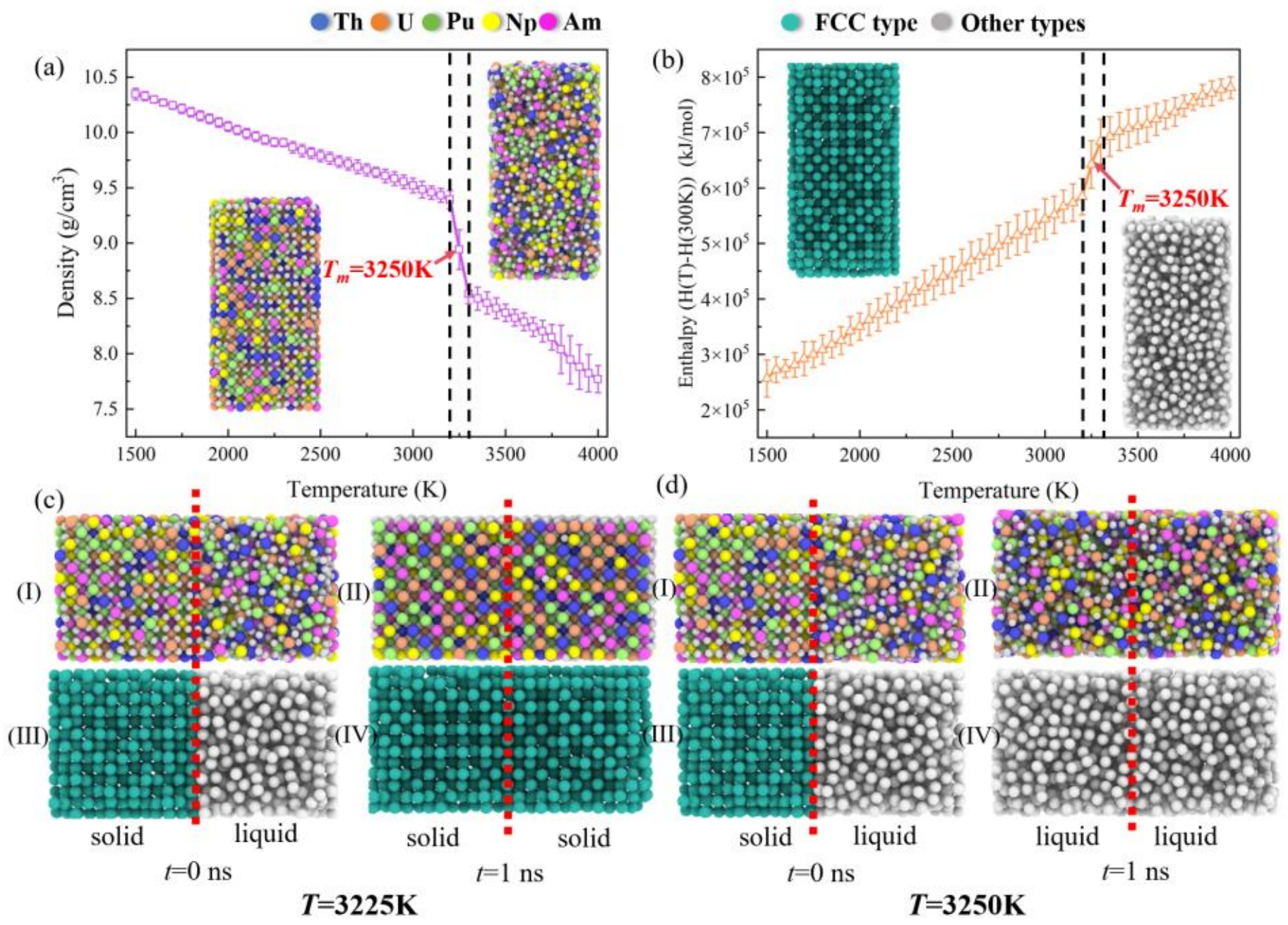

**Fig. 3** Two-phase simulation (TPS) measurements of (a) density and (b) enthalpy increment ( $H(T)-H(300\text{ K})$ ) variations for the HEC model across the temperature range of 1500–4000 K. Atomic snapshots at (c) $T = 3225$ K and (d) $T = 3250$ K illustrate the solid-liquid interface evolution. (I) Initial configuration containing coexisting solid and liquid phases at $t = 0$ ns. (II) Final configuration showing the complete transition into a pure solid ( $T = 3225$ K) or liquid ( $T = 3250$ K) phase at $t = 1$ ns. (III) and (IV) showing their corresponding Common Neighbor Analysis (CNA) results for (I) and (II), respectively, to distinguish between the FCC-structured solid and disordered liquid phases.

## 3 Results

### 3.1 Pair distribution function variations:

Understanding the structural evolution of the HEC at elevated temperatures is essential for its potential high-temperature applications. Fig. 4 illustrates the temperature-dependent variations of the pair distribution functions (PDFs). To clearly compare the distinct structural responses of the two sublattices, all cationic species are grouped into a single sublattice during the PDF evalutions. Notably, even at the upper temperature limit of 2800 K, the second peak of cationic PDF remains observable, indicating the crystalline integrity of the cationic sublattice (Fig. 4(a)). In contrast, the second peak of anionic PDF vanishes at 2600 K, signifying the emergence of a liquid-like state (Fig. 4(b)). This phenomenon demonstrates that anionic sublattice pre-melts at 2600 K. Fundamentally, anionic pre-melting signifies the onset of superionic transition, a second-order phase transition that significantly dictates the ionic diffusivity and thermophysical properties of the materials. Such transitions are prevalent in various fluorite and anti-fluorite materials, including $UO_2$ [46], MOX [14], and $Li_2O$ [47], typically occurring at a characteristic temperature of approximately 0.8-0.85 $T_m$. Therefore, the identification of superionic transition temperature ($T_S$) at 2600 K from the PDF plots is consistent with theoretical expectations, as its ratio to the HEC melting point ($T_m = 3250\,\text{K}$) closely approximates this empirical scaling law.

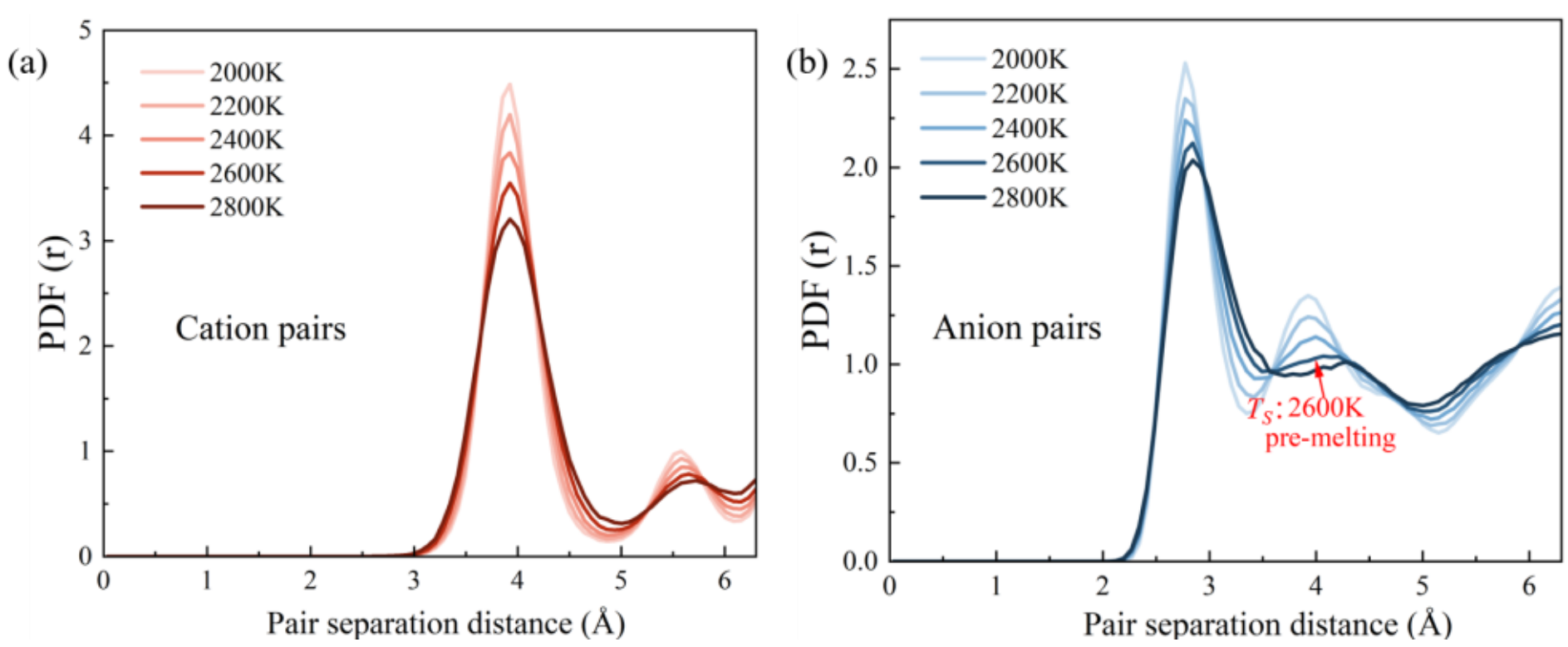

**Fig. 4** Variations in the pair distribution functions (PDFs) for (a) cationic and (b)

anionic pairs as a function of separation distance in the *M*2 HEC model at T= 2000–2800 K. $T_S$ denotes the superionic transition temperature of the HEC, corresponding to the pre-melting of anionic sublattice.

3.2 SRO-induced dynamic heterogeneity and confinement-enhanced diffusion

Traditional structural analysis methods often fall to characterize the highly disordered O-sublattice in the superionic state. To overcome this limitation, concepts originally developed for metallic glasses (MGs) [48] and supercooled glass-forming (GF) liquids [49] are adopted. Notably, the atomic entropy fingerprint developed by Piaggi and Parrinello [50] has proven highly effective in identifying local structural heterogeneities across various amorphous systems, including ceramics [51], MGs [52], and liquids [53]. This method captures local structural and thermodynamic features by coupling atomic configurational entropy with atomic enthalpy. The atomic entropy fingerprint for each atom $i$ is calculated via:

$$s^i = -2\pi\rho k_B \int_0^{r_m} \left[ g_m^i(r) \ln g_m^i(r) - g_m^i(r) + 1 \right] r^2 dr \qquad (1)$$

where $r$ is the distance between two atoms and $\rho$ is the number density. $g_m^i(r)$ is the modified radial distribution function of atom $i$, and it can be obtained via Eq. (2),

$$g_m^i(r) = \frac{1}{4\pi\rho r^2} \sum_j \frac{1}{\sqrt{2\pi\sigma^2}} e^{-(r-r_{ij})^2/(2\sigma^2)} \qquad (2)$$

where $\sigma$ denotes the width of Gaussians kernel used for smoothing $g(r)$, and $r_{ij}$ is the distance between the central atom $i$ and its neighbor $j$. The summation over $j$ incorporates all neighboring atoms within a cutoff radius $r_m$. Following the recommendations of the original methodology [50], the smoothing parameter $\sigma$ is set to 0.25, and the cutoff $r_m$ is defined as 7.7546 Å, which corresponds to 1.4 times the lattice constant of the HEC. (The calculated lattice constants of the SC HEC across the entire temperature range are detailed in Section S7 of the Supplementary Materials).

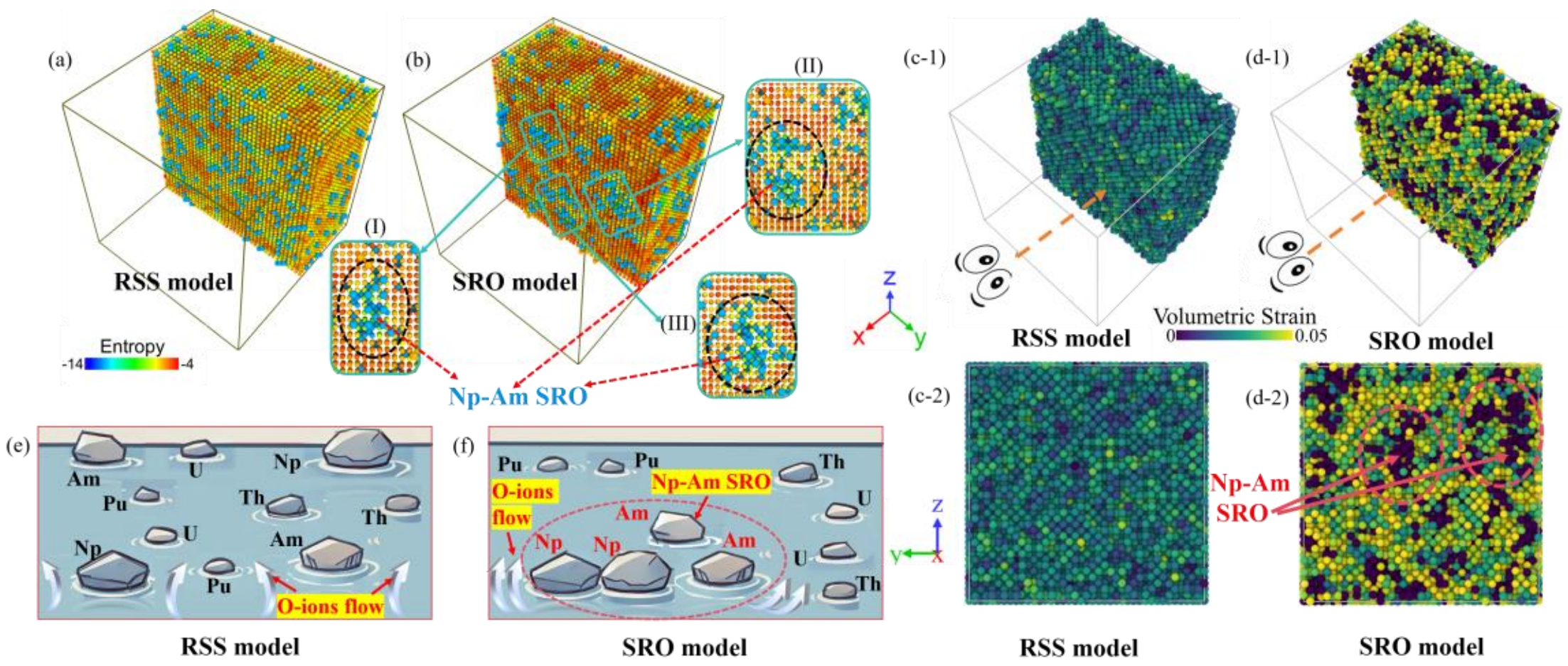

**Fig. 5.** Atomic entropy distributions for the (a) RSS and (b) SRO HEC models at 2600 K (with all ions colored to distinguish low-entropy solid cations from high-entropy liquid-like O ions), where zoom-in views (I) to (III) highlight the localized Np-Am SRO structures. Volumetric strain distributions of the cationic sublattices are shown for (c-1) RSS and (d-1) SRO models, accompanied by their respective cross-sectional views (c-2) and (d-2) taken along the indicated sightlines. These volumetric strain distributions explicitly present the lattice distortions induced by the Np-Am SRO, which constrain the random mobility of O-ions. Panels (e) and (f) provide schematic illustrations of superionic O-ion diffusion at 2600 K for both models. The light blue arrows delineate the presumed pathways of the O-ion "flow", illustrating

how the Np-Am SRO channels O-ion diffusion along preferential directions, ultimately giving rise to confinement-enhanced diffusion.

Fig. 5(a) and (b) present snapshots of the atomic entropy distributions in RSS and SRO HEC models at 2600 K. At this temperature, the anionic sublattice has entered a superionic state, characterized by the "liquid-like" diffusion of O ions [33, 54]. The color-coded entropy values visually display the distinct dynamic behaviors of the two sublattices: red indicates higher disorder and greater diffusion propensity of O ions, while blue corresponds to the low entropy of rigid cations. Unlike the uniform random distribution of cations in RSS model (Fig. 5(a)), Np and Am ions in SRO model (Fig. 5(b)) aggregate to form Np-Am SRO structures. As highlighted in the zoom-in views (I) to (III), these localized SRO domains are enclosed by lower-entropy O ions (green and yellow), whereas the highly mobile, high-entropy O ions (red) are distributed further away. This spatial arrangement suggests that the Np-Am SRO obstructs the local diffusion of superionic O ions, effectively directing the anionic transport toward kinetically favorable regions. Additionally, as evidenced by comparing Fig. 5(a) and (b), the SRO model exhibits a significantly higher overall anionic entropy than RSS model. Such elevated entropy indicates that anionic mobility is concurrently enhanced within the unblocked channels of SRO model, leading to a phenomenon defined as a confinement-enhanced diffusion effect. Fundamentally, this confinement effect originates from the severe lattice distortions induced by the Np-Am SRO. The localized Coulomb repulsion within these SRO regions generates a local electrostatic and strain-coupled potential well, triggering a blocking effect that restricts adjacent anionic motion [33, 55]. This mechanism is strongly corroborated by the volumetric strain distributions (Fig. 5(c-1) and (d-1)). Detailed derivations of volumetric strain are provided in Section S8 of the Supplementary Materials. Crucially, the localized strain fields spatially correlate with the low-entropy regions, confirming that the biased distribution of mobile O ions is dictated by the SRO-induced lattice distortion. This spontaneous kinetic rearrangement manifests as pronounced dynamic heterogeneity within the oxygen sublattice. To intuitively conceptualize this phenomenon, the superionic phase can be viewed as a fluid-like environment (Fig. 5(e) and (f)), where the highly mobile O ions “flow” through the stationary cationic skeleton. In the RSS model, the O-ions “flow” randomly. Conversely, the Np-Am SRO structure in the SRO model act as rigid obstacles, generating strain fields that block the O-ions “flow” routes. This blocking effect forces the O-ions “flow” to bypass the SRO regions and form concentrated transport pathways, thereby activating the onset of dynamic heterogeneity observed in the superionic HEC system.

### 3.3 Quantitative assessment of SRO-driven dynamic heterogeneity

To quantitatively analyze the spatial heterogeneity of O-ions discussed in Fig. 5, we adopt the nonaffine squared displacement ($D^2_{\min}$) as an indicator of local mobility. Fundamentally, $D^2_{\min}$ effectively captures the irreversible atomic displacements associated with local structural rearrangements and the loss of centrosymmetry, making it a robust metric for identifying active diffusion sites in highly disordered or amorphous systems [56, 57]. Fig. 6(a) and 6(b) provide snapshots of the $D^2_{\min}$ spatial distribution in O-sublattice for RSS and SRO models across various temperatures. At 2200 and 2400 K, both models exhibit similar mobility profiles, although regions with elevated $D^2_{\min}$ values (colored in red) appear slightly more clustered in the SRO model. However, distinct spatial disparities emerge above the superionic transition temperature ($T_S = 2600$ K). In the SRO model, highly mobile O ions with larger $D^2_{\min}$ strongly aggregate, forming spatially interconnected, channel-like active regions. This observation aligns with the heterogeneous mobility patterns reported in glass-forming liquids, characterized by highly active “hot channels” surrounded by localized, low-mobility pockets [49]. Furthermore, at temperatures ranging from 2600 to 3000 K, the SRO model exhibits a higher overall density of these active diffusion sites compared to the RSS model, directly corroborating the confinement-enhanced mobility mechanism observed in Section 3.2.

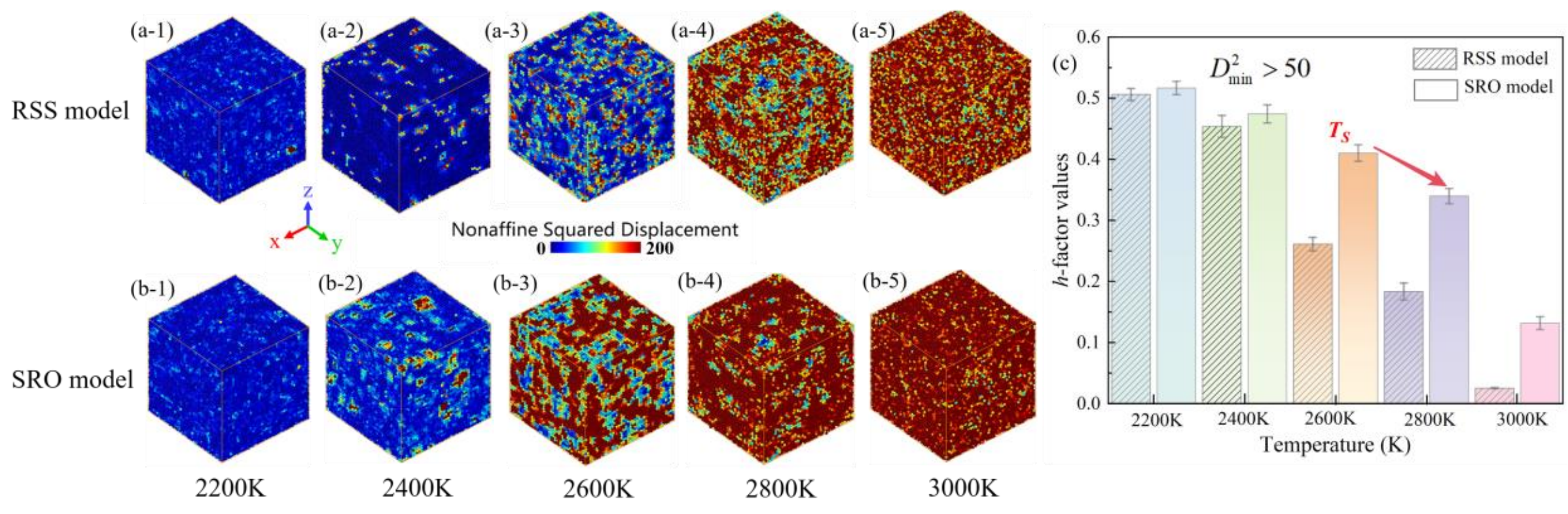

**Fig. 6** Quantitative analysis of the dynamic heterogeneity in the O-sublattice.

Atomic snapshots displaying the spatial distribution of nonaffine squared displacement ($D^2_{\min}$) in the (a) RSS and (b) SRO HEC models from 2200 to 3000

K. The O ions are colored according to their $D^2_{\min}$ values. (c) Variations of the spatial inhomogeneity parameter (*h*-factor) as temperature increases. The *h*-factor are evaluated exclusively for highly mobile O ions ($D^2_{\min} > 50$). Shaded and solid bars denote the RSS and SRO models, respectively. Each bar represents the average *h*-factor calculated across ten configurations during a 1 ns equilibrium trajectory at the given temperature, with error bars indicating the standard deviation of mean.

To statistically quantify the spatial heterogeneity of anionic mobility, we employ the *h*-factor for the mobile O-ions (defined here as those with $D^2_{min} > 50$), as presented in Fig. 6(c). Originally developed for network topology [58] and subsequently adapted to characterize stress-induced spatial inhomogeneity in amorphous materials [59], the *h*-factor serves as a metric ranging from 0 (completely uniform) to 1 (highly clustered). Detailed information of the *h*-factor is provided in Section S9 of the Supplementary Materials. As illustrated in Fig. 6(c), the *h* -factor values for both models generally decrease from 2200 to 3000 K, suggesting that elevated thermal activation gradually homogenizes the spatial distribution of mobile O ions. Although the disparity between the two models remains marginal from 2200 to 2400 K, a remarkable divergence emerges as the system reaches the superionic transition temperature ($T_S = 2600$ K). At 2600 K, the *h*-factor of RSS model experience remarkably drop, indicating that the anionic sublattice is overwhelmed by thermal fluctuations and approaches a homogeneous fluid-like state. Conversely, the SRO model maintains a significantly higher *h*-factor. This signals that the blocking effect imposed by the Np-Am SRO resists thermal homogenization by establishing preferential, directional pathways for anionic diffusion. This structural adaptation implies that mobile anions are forced into interconnected transport networks [24, 27], whose underlying topological origin will be discussed in the subsequent section. Eventually, upon reaching 3000 K, the intense thermal disorder approaching melting point completely overrides the orienting influence of the SRO, causing the dynamic heterogeneity in both models to collapse.

3.4 Topological origin of dynamic self-organization and cooperative diffusion

To characterize the highly disordered topology of superionic O-sublattice, we performed atomic cluster analysis using Voronoi tessellation and coordination polyhedral analysis method [60]. The anionic Voronoi polyhedra are represented by the index $<n_3, n_4, n_5, n_6>$, where $n_i$ denotes the number of *i*-edged faces. Fig. 7(a) and (b) illustrate the fractions of the most abundant Voronoi polyhedra in HEC SRO model across various temperatures. Notably, the anionic polyhedra in HEC RSS model exhibit nearly identical proportions and changing trends (see Section S10 of the Supplementary Materials). Below $T_S$ (2000-2400 K), anionic sublattice is predominantly composed of <1,3,3,1>, <2,3,2,1>, and <2,2,2,2> polyhedra. However, as the temperature approaches $T_S$ at 2600 K, the populations of these intermediate-temperature polyhedra decline rapidly. Concurrently, the proportion of the <0,4,4,0>-polyhedron increases, ultimately dominating the anionic sublattice in the superionic state (Fig. 7(a) and (b)). Furthermore, the face distribution (Fig. 7(c)) demonstrates a persistent dominance of 4- and 5-edged faces at elevated temperatures, indicating the prevalent formation of five-fold local symmetry and topological frustration, which are universal structural hallmarks of liquid-like mobility [61].

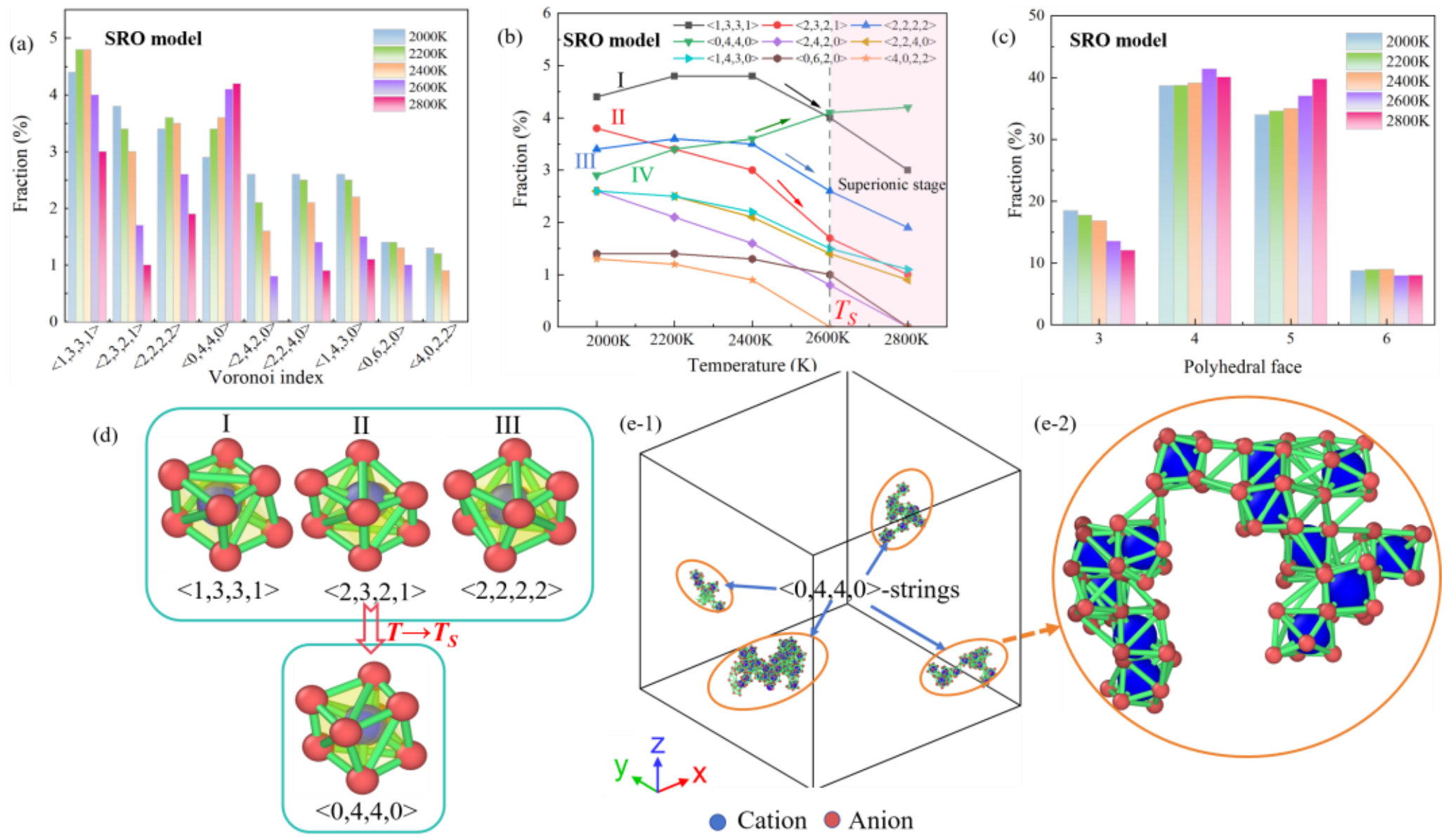

**Fig. 7** Topological evolution of the O-sublattice in HEC SRO model via Voronoi cluster analysis. (a) Fractions of the nine most abundant anionic Voronoi polyhedra at temperatures ranging from 2000 to 2800 K. (b) Temperature-dependent variations of the dominant Voronoi polyhedral fractions. The dashed line marks the onset of superionic transition at $T_S$. (c) Distribution of polyhedral faces (3-, 4-, 5-, and 6-edged) fractions across different temperatures. (d) Structural snapshots illustrating the topological change from intermediate-temperature polyhedra (I: <1,3,3,1>, II: <2,3,2,1>, III: <2,2,2,2>) to high-temperature <0,4,4,0> dominant polyhedron at $T_S$. (e-1) Spatial distribution and (e-2) enlarged view of the spatially interconnected <0,4,4,0> string-like polyhedral networks at 2600 K, which act as continuous transport pathways for the superionic O-ions.

Crucially, the emergence of <0,4,4,0> cluster governs the dynamic self-organization of O-ions. Rather than remaining isolated, these highly dense <0,4,4,0>-polyhedra assemble into continuous, spatially interconnected networks as their proportion increases (Fig. 7(e-1) and (e-2)). This spatially heterogeneous agglomeration gives rise to cooperatively rearranging regions (CRRs) [62]. Consequently, these densely packed, string-like <0,4,4,0> networks act as the structural carriers for the collective motion of superionic O-ions (visually confirmed in Section S11 of the Supplementary Materials). Interestingly, similar hexatic phases emerge during the solid-to-liquid transitions of nanoconfined water [63], underscoring the physical universality of dominant polyhedral networks in superionic or fluid-like states. In essence, the proliferation of <0,4,4,0>-polyhedron weaves preferentially oriented fast-diffusion networks within the superionic HEC, elucidating the topological origin of the dynamic heterogeneity and confinement-enhanced transport observed in Figs. 5 and 6.

## 4 Discussion

### 4.1 Thermodynamic driving forces for the superionic topological networks

To unveil the thermodynamic driving forces behind the self-organization of the <0,4,4,0>-polyhedron, we quantify the configurational entropy of anionic sublattice based on the topological diversity of the Voronoi polyhedra. Using the Shannon entropy model, the statistical complexity of topological configurations can be calculated as:

$$S_{Shannon} = -\sum_{i=1}^{n} p_i(x_i)\log(p_i(x_i)) \tag{3}$$

By introducing the Boltzmann constant ($k_B$), the Gibbs entropy ($S_{Gibbs}$) of Voronoi

polyhedra is defined as:

$$S_{Gibbs} = -k_B \sum_{i=1}^{n} p_i(x_i)\log(p_i(x_i)) \tag{4}$$

where $p_i(x_i)$ denotes the probability distribution of a specific Voronoi face. Subsequently, the Gibbs free energy ($G$) of the system is determined by the standard thermodynamic relation:

$$G = H - TS_{Gibbs} \tag{5}$$

Here, the enthalpy ($H$) is estimated to be -13.537 eV, aligning with experimental reports [64]. Given the severe lattice distortions inherent to the HEC, configurational entropy overwhelmingly dominates its vibrational counterpart [65]. Thus, the vibrational contribution can be neglected in this derivation.

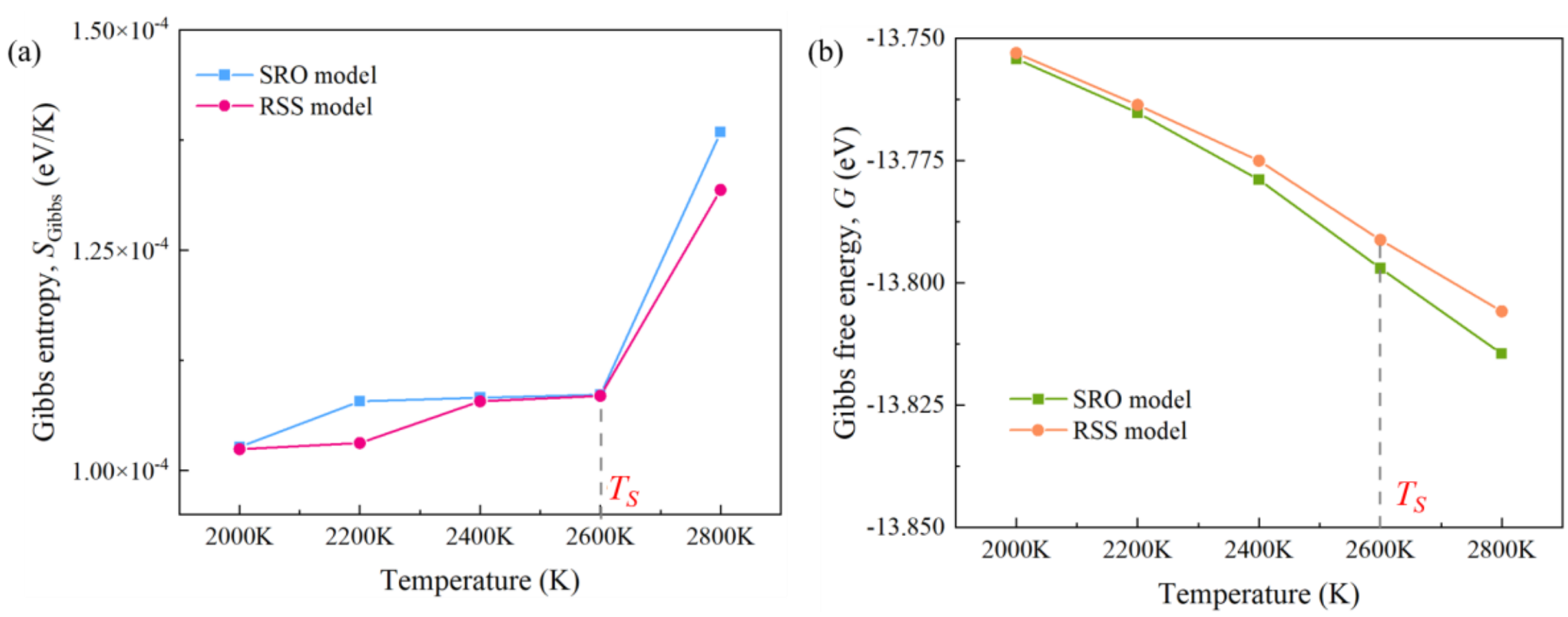

**Fig. 8** Variations of the (a) Gibbs entropy ($S_{Gibbs}$) and (b) Gibbs free energy ($G$) as a function of temperature for the RSS and SRO HEC models.

Fig. 8 illustrates the temperature-dependent variations of the Gibbs entropy ($S_{Gibbs}$) and Gibbs free energy ($G$). As the temperature increases, $S_{Gibbs}$ increases significantly at the $T_S$ of 2600 K, signaling a structural transformation toward a highly disordered state. Concurrently, $G$ decreases monotonically across the $T_S$. Such results indicates the system spontaneously adopts these highly distorted <0,4,4,0>-polyhedral networks as transport pathways to maximize its configurational entropy during the superionic transition. Ultimately, this topological adaptation constitutes an energetically favorable configuration that stabilizes the HEC by effectively minimizing its total free energy at elevated temperatures. Driven by these thermodynamic imperatives, the system spontaneously adopts the highly distorted <0,4,4,0>-polyhedral networks as topological transport pathways to maximize its configurational entropy. Ultimately, this structural adaptation constitutes an energetically favorable configuration, stabilizing the HEC by minimizing its total free energy at elevated temperatures. Furthermore, the SRO model consistently maintains a higher $S_{Gibbs}$ (Fig. 8(a)) and a lower $G$ (Fig. 8(b)) than RSS model across the entire temperature range, particularly within the superionic regime. Consequently, it is this superior energetic favorability of SRO model that provides the thermodynamic driving force for the formation of Np-Am short-range order.

### 4.2 Transient caging, structural relaxation, and anomalous kinetic inversion

To translate this thermodynamic favorability into kinetic performance, we quantitatively evaluate the impact of the SRO and superionic transition on diffusion kinetics using the mean squared displacement (MSD) of all ions, calculated via:

$$MSD(t) = \frac{1}{N}\sum_{i=1}^{N}\langle |\mathbf{r}_i(t) - \mathbf{r}_i(0)|^2 \rangle \tag{6}$$

where $N$ is the total number of mobile species, $\mathbf{r}_i(t)$ and $\mathbf{r}_i(0)$ denote the current and reference position vectors of ion $i$, respectively. To comprehensively accelerate the dynamic processes without altering the intrinsic mechanisms, an isotropic hydrostatic stress of 100 MPa is applied. Following a 1 ns equilibration under the NPT ensemble, atomic trajectories are recorded over a 100 ps run to derive the MSDs.

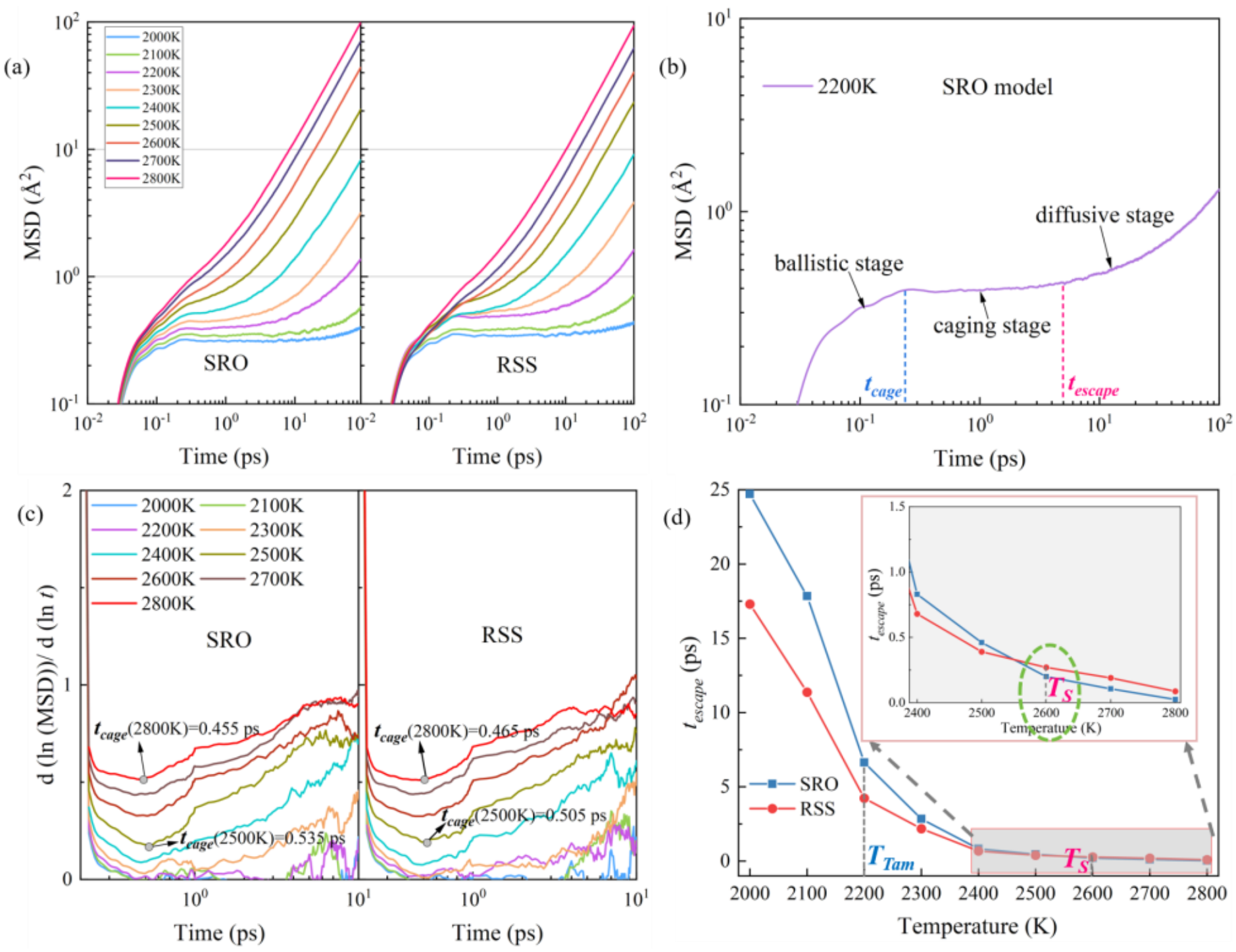

**Fig. 9**. (a) Mean squared displacement (MSD) of O-ions in SRO and RSS HEC models at various temperatures. (b) Representative MSD trajectory in the SRO model at 2200 K, delineating the ballistic, caging, and diffusive stages, demarcated by the cage formation time ($t_{cage}$) and cage escape time ($t_{escape}$). (c) Logarithmic derivative of the MSD versus time for SRO and RSS models, the local minima correspond to $t_{cage}$. (d) Temperature dependence of $t_{escape}$ for SRO and RSS models. The disparity in $t_{escape}$ between the two models decreases significantly as approaches the Tammann crossover temperature ($T_{Tam}$). The inset highlights a kinetic inversion, where the SRO model achieves a shorter $t_{escape}$

than RSS model upon crossing the superionic transition temperature ($T_S$).

As illustrated in Fig. 9(a), the MSD curves of O-ions reveal distinct temperature-dependent diffusion regimes. Taking the SRO model at 2200 K as a representative example (Fig. 9(b)), the early timescale exhibits a quadratic growth characteristic of the "ballistic stage," where ions move freely before interacting with their nearest neighbors. As time progresses, the MSD curve shifts to a plateau, known as the "caging stage," where ions are temporarily confined by their surrounding atomic framework. The onset of this localized trapping is quantified by the cage formation time ($t_{cage}$), which is determined by the local minima in the logarithmic derivative of the MSD (Fig. 9(c)) [66]. Below $T_S$, the $t_{cage}$ in SRO model is visibly greater than that in RSS model (e.g., 0.535 ps vs. 0.505 ps at 2500 K), confirming that the Np-Am SRO structures exert a more rigid localized caging effect on O-ions. However, to initiate long-range diffusion, these trapped ions must break free from their local coordination polyhedra. This critical duration, known as the cage escape time ($t_{escape}$), effectively represents the structural relaxation time of the anionic sublattice (analogous to the α-relaxation time in glassy materials) [33]. As displayed in Fig. 9(d), $t_{escape}$ decreases with increasing temperature in both models due to enhanced thermal activation. Notably, the difference in $t_{escape}$ between the RSS and SRO models drops significantly near a critical temperature of 2200 K. This threshold corresponds to the so-called Tammann crossover temperature ($T_{Tam}$) [67], which marks a pronounced increase in ionic diffusion. Unlike the behavior at elevated temperatures, the relative magnitude of $t_{escape}$ between two models does not reverse at $T_{Tam}$, indicating that the fundamental diffusion mechanism remains unaltered in this intermediate regime. Crucially, a striking kinetic inversion emerges at the superionic threshold ($T_S = 2600$ K). While the SRO model restricts anionic escape at lower temperatures (yielding a longer $t_{escape}$), it facilitates a significantly faster escape than the RSS model once the superionic threshold ($T_S$) is crossed. Crucially, this kinetic inversion serves as a direct dynamic reflection of the topological characteristics of SRO atomic arrangement. The Np-Am SRO structures, which act as the structural traps at lower temperatures, undergo a transformation at $T_S$. By spontaneously weaving highly interconnected <0,4,4,0>-polyhedral topological networks, they facilitate highly efficient, collective cage-escaping motions, ultimately triggering the anomalous diffusion enhancement observed in the superionic state. Once the time exceeds $t_{escape}$, the O-ion dynamics transits into the diffusive stage, where the MSD can be utilized to extract long-time transport properties.

## 4.2 Non-Arrhenius diffusion and determination of the Tammann temperature

To extract the diffusion coefficients ($D(O)$) and activation energies ($Q_d$), the MSD data are fitted using the Einstein relation and Arrhenius equation:

$$D = \lim_{t \to \infty} \frac{MSD(t)}{6t} \tag{7}$$

$$D(O) = D_0 \exp\left(-\frac{Q_d}{k_B T}\right) \tag{8}$$

where $D_0$ represents the pre-exponential factor, $k_B$ is the Boltzmann constant, and $T$ denotes the absolute temperature.

Fig. 10(a) presents the $\ln D(O)$ vs. $1/k_B T$ plots for RSS and SRO HEC models. Clearly, the O-ions in both models exhibit a pronounced deviation from the typical

Arrhenius relation. This anionic "non-Arrhenius" diffusion is widely reported in materials undergoing the superionic transition (e.g., $UO_2$, MOX) [34, 45, 68]. We utilize the aforementioned Tammann crossover (2200 K) and superionic transition (2600 K) temperatures to divide the overall curves into three linear segments with distinct slopes, corresponding to varying diffusion activation energies. Notably, the activation energies in the mid-temperature (MT) and low-temperature (LT) regimes are nearly identical. This invariant activation energy indicates the persistence of underlying anionic diffusion mechanism across the Tammann temperature, agreeing with the structural relaxation results observed in Fig. 9. However, the high-temperature (HT, $T > T_S$) activation energies for both models ($Q_{d,RSS}^{HT} = 2.462 \pm 0.133$ and $Q_{d,SRO}^{HT} = 2.964 \pm 0.375$) are significantly lower than those observed at the MT and LT stages ($Q_{\mathrm{d},RSS}^{MT,LT} = 4.732 \pm 0.004$ and $Q_{\mathrm{d},SRO}^{MT,LT} = 4.855 \pm 0.632$). This substantial reduction signifies a profoundly enhanced O-ion mobility above $T_S$, further demonstrating that the Tammann crossover exerts a weaker impact on altering the anionic diffusion mechanism compared to the superionic transition. Such a result is physically expected, as the O-ion transport mechanism shifts from localized site-hopping to liquid-like collective diffusion in the superionic phase [16, 33]. Remarkably, while the HT activation energy of the SRO model (2.964 eV) remains higher than that of RSS model (2.462 eV), its diffusivity ultimately surpasses the RSS model at 2800 K. This kinetic paradox can be interpreted by the Meyer-Neldel compensation rule. According to transition state theory, the pre-exponential factor $D_0$ correlates with the kinetic activation entropy ($\Delta S_{act}$) and the multiplicity of available diffusion pathways, scaling mathematically as $D_0 \propto \exp(\Delta S_{act} / k_B)$ [69]. Although the Np-Am SRO imposes a slightly higher energetic barrier for O-ion diffusion, the self-organization of <0,4,4,0>-polyhedron weaves a dense, highly interconnected transport network. This topological adaptation, which maximizes the system's configurational entropy as proved in Section 4.1, translates into a massive increase in kinetic activation entropy, thereby exponentially elevating the pre-exponential factor ( $D_0$ ). At elevated temperatures within the superionic regime, this explosive increase in pathway availability completely overcompensates for the slightly higher energy barrier ( $Q_d$ ). Consequently, this entropy-driven compensation enables the confinement-enhanced diffusion observed in the SRO model, allowing its high-temperature diffusivity to overtake that of RSS model.

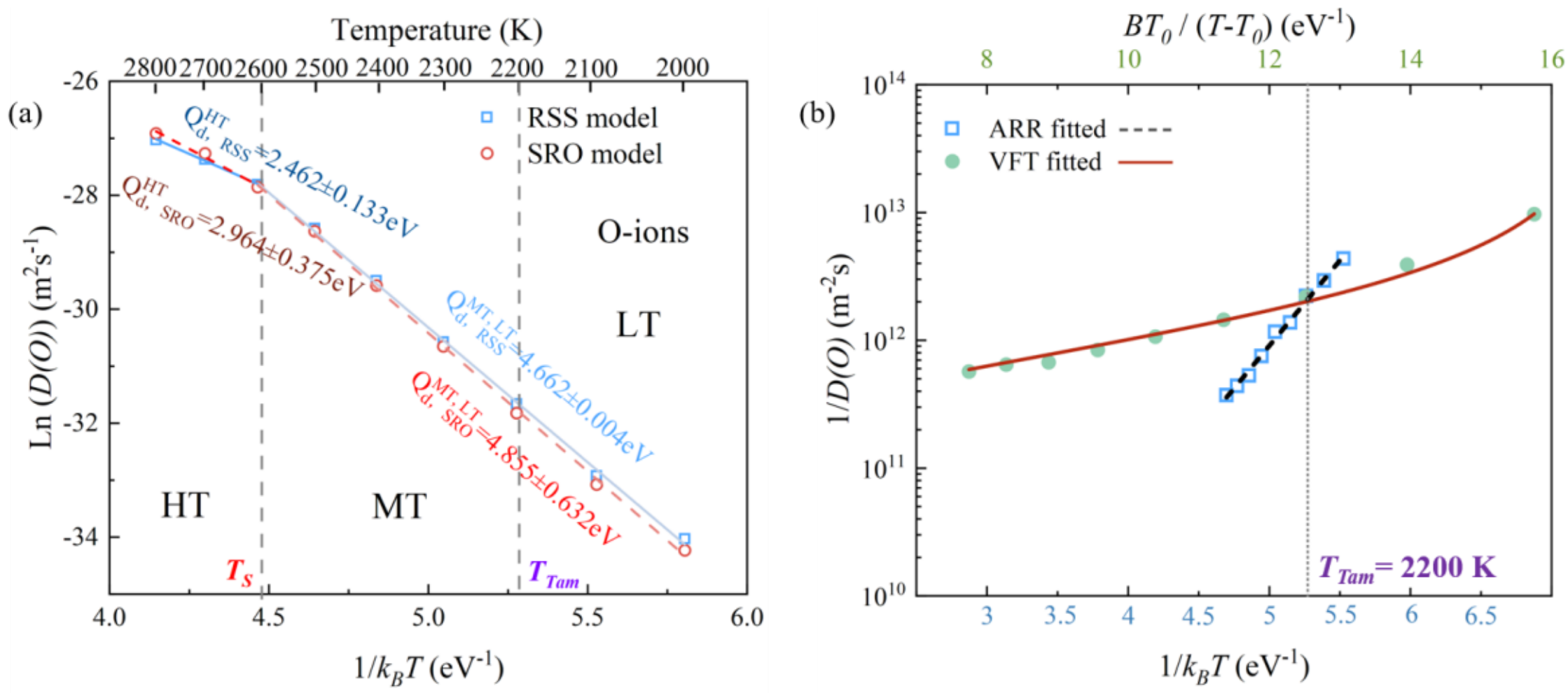

**Fig. 10.** (a) Arrhenius plots of the O-ion diffusion coefficients ( $D(O)$ ) for the RSS and SRO HEC models. The entire temperature range is segmented into three regimes (HT, MT, and LT) demarcated by the superionic transition temperature ( $T_S = 2600$ 2600 K) and the Tammann temperature ( $T_{Tam} = 2200\,\text{K}$ ). These regimes yield distinct

activation energies corresponding to high-temperature ($Q_{d,SRO}^{HT}$, $Q_{d,RSS}^{HT}$) and the mid- to low-temperature ($Q_{d,SRO}^{MT,LT}$, $Q_{d,RSS}^{MT,LT}$) stages. (b) Variations of the reciprocal anionic diffusivity ($1/D(O)$) versus $1/k_BT$ in SRO HEC model. The solid and dashed lines represent fitting via the Vogel-Fulcher-Tammann (VFT) and Arrhenius (ARR) equations, respectively. The precise intersection of these two models analytically determines the Tammann temperature ($T_{Tam}$) at approximately 2200 K.

While $T_S$ defines the structural transition to the superionic phase, the dynamical onset of cooperative mobility originates at a lower temperature, characterized by the Tammann crossover temperature ($T_{Tam}$) [33]. Drawing an analogy between the Tammann crossover in superionic materials and glass transition in amorphous materials, the ionic transport exhibits a dynamic shift from "strong" to "fragile" liquid behavior [70]. Consequently, the low- and high-temperature mobilities are typically described by the Arrhenius (ARR) and Vogel-Fulcher-Tammann (VFT) equations, respectively. Unlike the purely thermally activated ARR model, the VFT equation accounts for cooperative atomic rearrangements and local caging effects [24, 33], expressed as [71]:

$$D_{VFT} = D_0 \exp\left(\frac{-B_{VFT} \cdot T_0}{T - T_0}\right) \tag{9}$$

where $B_{VFT}$ serves as a measure of the system's fragility (related with the activation barrier), and $T_0$ denotes the “Vogel temperature” or “ideal glass transition temperature”. Conversely, the ARR equation assume that the diffusion process is predominated by a thermal activation mechanism:

$$D_{ARR} = D_0 \exp\left(\frac{-E_a}{k_B T}\right) \tag{10}$$

where $E_a$ is the diffusion activation energy. As established in glass-forming liquids, the transition from Arrhenius to VFT relaxation occurs at the glass transition temperature ($T_g$) [72, 73]. A physically analogous transition is observed in superionic-state species at their Tammann temperature ($T_{Tam}$), marking the exact crossover point where ion diffusion transforms into a highly cooperative process [33]. Hence, $T_{Tam}$ can be analytically determined by mathematically equating the VFT and ARR equations ($D_{VFT} = D_{ARR}$), yielding:

$$\frac{B_{VFT} \cdot T_0}{T_{Tam} - T_0} = \frac{E_a}{k_B \cdot T_{Tam}} \tag{11}$$

By solving this equivalence, the exact expression for the Tammann temperature ($T_{Tam}$) is derived as:

$$T_{Tam} = \frac{E_a \cdot T_0}{E_a - B_{VFT} \cdot k_B \cdot T_0} \tag{12}$$

Considering that the HEC model exhibits thermodynamic similarities to $UO_2$ (e.g., melting point $T_m$, superionic transition temperature $T_S$, and lattice constants), the empirical fragility parameters characterizing the VFT behavior are reasonably adopted from established $UO_2$ [70]. By determining the precise intersection between the VFT model and Arrhenius model, we can theoretically pinpoint this crossover point. As evidenced by the reciprocal diffusivity curves in Fig. 10(b), this analytical approach yields a Tammann crossover temperature ($T_{Tam}$) of 2200 K for the HEC, agreeing perfectly with that inferred from the structural relaxation times in Fig. 9.

### 4.4 Effective viscosity of the liquid-like anionic sublattice

To further validate the "liquid-like" nature of superionic anionic sublattice and to cross-verify the kinetic inversion observed in the Arrhenius plots (Fig. 10), we examine the flow resistance of the O-ions in terms of viscosity. According to classical fluid dynamics, the effective viscosity ($\eta$) of these highly mobile anions can be estimated from their diffusivity via the Stokes-Einstein relation [74, 75]:

$$D(O) = \frac{k_B T}{6\pi\eta r_i} \tag{13}$$

where $r_i$ denotes the effective ionic radius of the O-ions. The ionic radius can be accurately extracted by determining the local coordination number (CN) and cross-referencing Shannon's tabulated empirical radii [76]. The CN is quantitatively integrated from the pair distribution function (PDF) via:

$$CN = 4\pi\rho \int_0^{r_{\min}} r^2 g(r) dr \tag{14}$$

where $\rho$ represents the atomic number density, $g(r)$ denotes the PDF profile (as provided in Fig. 4b), and $r_{\min}$ is the position of first minimum. Using the extracted ionic radius ($r_i = 1.38$ Å) and previously established scaled diffusivity ($D_O / T$, Fig. 10b), the temperature-dependent effective viscosity of anions is calculated by,

$$\eta = \frac{k_B}{6\pi r_i} \cdot \frac{T}{D(O)} \tag{15}$$

To facilitate a precise comparison of the flow resistance between the SRO and RSS HEC models across a wide temperature range, we define a normalized viscosity ($\eta^*$) relative to a trapped, low-temperature reference state (2000 K):

$$\eta^* = \frac{\eta(T)}{\eta(2000K)} \tag{16}$$

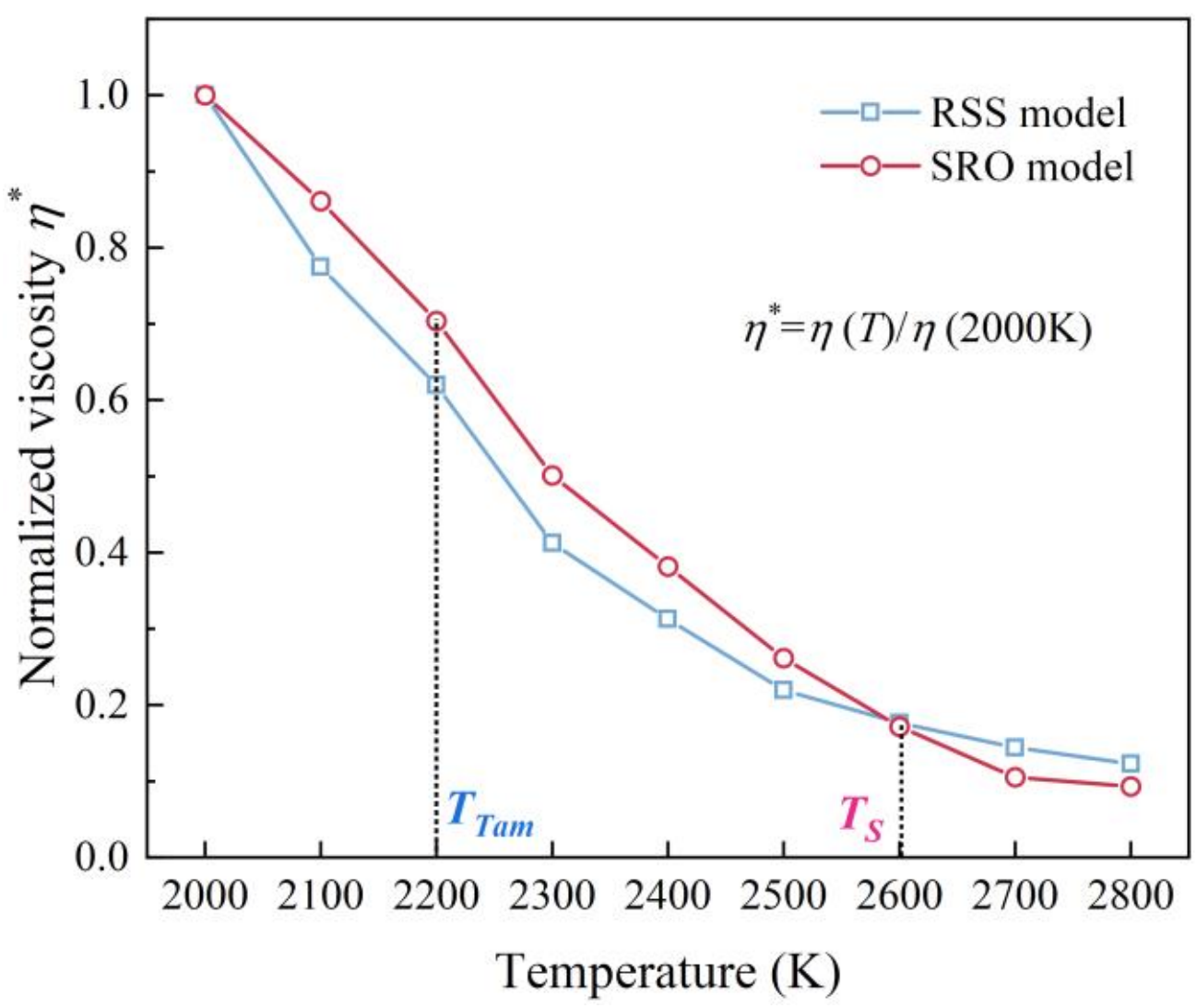

**Fig. 12.** Temperature dependence of the normalized viscosity ($\eta^*$) for anionic sublattice in RSS and SRO HEC models. The normalized parameter $\eta^*$ is defined as the ratio of effective viscosity at a given temperature to the reference viscosity at 2000 K ($\eta^* = \eta(T)/\eta(2000\text{ K})$). The critical boundaries are demarcated by the Tammann crossover temperature ($T_{Tam} = 2200$ K) and superionic transition temperature ($T_S = 2600$ K). The plot highlights a kinetic inversion, where the viscosity of SRO model drops below that of RSS model upon surpassing $T_S$.

As shown in Fig. 12, the normalized viscosity ($\eta^*$) exhibits a monotonic decrease with increasing temperature, indicating the progressive fluidization of the anionic sublattice. At lower temperatures, the SRO model maintains a slightly higher $\eta^*$ than RSS model, confirming that the rigid Np-Am SRO exert substantial viscous drag against anionic diffusion. However, analogous to the critical deflection observed in Arrhenius plots, the $\eta^*$ of SRO model drops below that of RSS model upon surpassing 2600 K, signaling the onset of kinetic inversion at $T_S$. This viscosity inversion serves as proof that the SRO-induced <0,4,4,0>-polyhedral network transform localized structural traps into a highly lubricated transport highway, thereby enabling enhanced, liquid-like O-ion mobility. Furthermore, the precise alignment of this viscosity inversion at $T_S$, coupled with the absence of any transition occurs near $T_{Tam}$, conclusively demonstrates that the anomalous enhancement in O-ion diffusion is governed by the structural superionic transition rather than dynamical Tammann crossover.

### 4.5 Dynamical heterogeneity of the O-ion diffusion

Despite previous analyses demonstrating that the shift in anionic diffusion mechanism is governed by superionic transition at $T_S$, VFT kinetics and the emergence of Tammann crossover ($T_{Tam}$) imply that O-ion transport is underpinned by cooperative, spatially heterogeneous dynamics. To quantitatively evaluate this dynamical heterogeneity, the non-Gaussian parameter ($\alpha_2$) of O-ions is calculated [77]. This parameter tracks the deviation of ionic dynamics from a random-walk Gaussian distribution over a given time interval, defined as:

$$\alpha_2(\Delta t) = \frac{3\langle \Delta r^4(\Delta t)\rangle}{5\langle \Delta r^2(\Delta t)\rangle^2} - 1 \tag{17}$$

where $\Delta r(\Delta t)$ denotes the displacement distance of an individual ion during the time interval $\Delta t$, and angle brackets $\langle \ldots \rangle$ represent the ensemble average over all O-ions.

Accordingly, an $\alpha_2$ value of 0 signifies pure Gaussian-type random-walk diffusion.

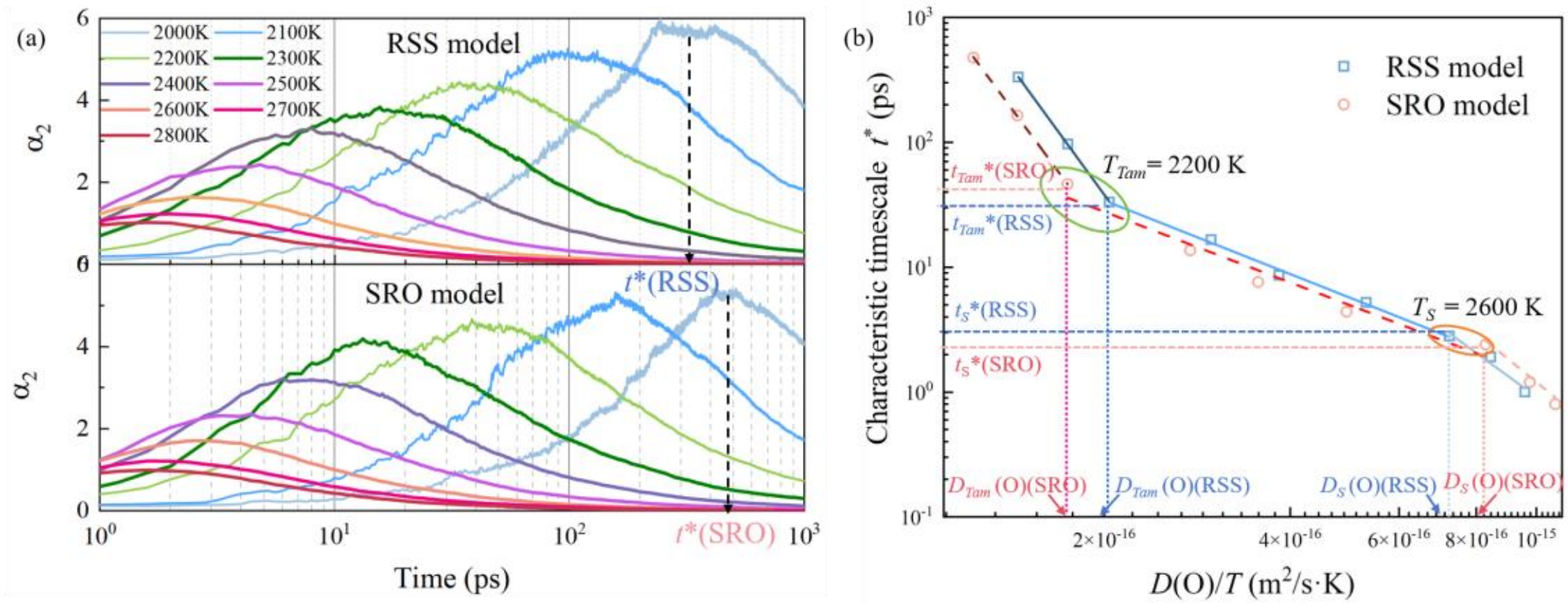

**Fig. 11** Quantification of the dynamical heterogeneity for O-ions. (a) Time evolution of the non-Gaussian parameter ($\alpha_2$) for O-ions in RSS and SRO models across varying temperatures. (b) Correlation between the characteristic timescale of dynamic heterogeneity ($t^*$, extracted from the $\alpha_2$ peaks) and the scaled diffusivity ($D(O)/T$). The distinct linear regimes are divided by the Tammann crossover temperature ($T_{Tam} = 2200$ K) and superionic transition temperature ($T_S = 2600$ K). $t^*_{Tam}(RSS, SRO)$ and $t^*_S(RSS, SRO)$ denote the characteristic timescales at these critical temperatures for both models, while $D_{Tam}(O)(RSS, SRO)$ and $D_S(O)(RSS, SRO)$ represent their corresponding diffusion coefficients.

As shown in Fig. 11(a), the diffusion of O-ions in both RSS and SRO models exhibits typical non-Gaussian behavior, with $\alpha_2$ reaching a prominent peak at a specific time. Physically, this peak represents the exact moment when the spatial disparity between fast-moving ("liquid-like") O-ions and slow-moving (trapped) O-ions is maximized within the system; this critical point defines the characteristic timescale of dynamic heterogeneity ( $t^*$ ). The peak magnitude of $\alpha_2$ decreases monotonically as the temperature increases, consistent with behavior widely observed in glass-forming liquids and other superionic crystals exhibiting dynamical heterogeneity [33]. To further elucidate the interrelation between this characteristic timescale ($t^*$) and O-ion mobility, we correlate $t^*$ with the scaled diffusivity ($D(O)/T$). As plotted in Fig. 11(b), distinct scaling relationships between $t^*$ and $D(O)/T$ emerge, distinctly segmented by the two critical temperatures, $T_{Tam}$ and $T_S$ . Notably, in the low-temperature regime (below $T_{Tam}$), the SRO model exhibits a longer characteristic timescale ($t^*_{Tam}(SRO) > t^*_{Tam}(RSS)$) and lower diffusivity compared to the RSS model ( $D_{Tam}(O)(SRO) < D_{Tam}(O)(RSS)$ ) at the same temperature. However, within the superionic regime (above $T_S$), this kinetic trend completely reverses: the SRO model achieves a remarkably shorter timescale ($t^*_S(SRO) < t^*_S(RSS)$) and higher diffusivity than the RSS model ($D_S(O)(SRO) > D_S(O)(RSS)$). This kinetic inversion indicates a mechanistic transition triggered by the specific atomic arrangement: the Np-Am SRO structures act as rigid structural traps that severely hinder ionic mobility at lower temperatures, yet they transform into topological facilitators that promote highly efficient, collective ion migration once the superionic threshold is breached.

More importantly, this segmented scaling behavior delineates the distinct physical roles of these two critical transitions. While the superionic transition ($T_S$) dictates the shift in anionic diffusion mechanism and ultimately triggers the kinetic inversion, the Tammann crossover ($T_{Tam}$) serves as a precursor that modulates the extent of dynamical heterogeneity prior to the superionic transition.

4.6 Defect proliferation and structural breakdown of the anionic sublattice

Having established that the superionic transition ($T_S$) activates the kinetic inversion and Tammann crossover ($T_{Tam}$) modulates dynamical heterogeneity, we now shift our focus to the structural origins of these transport phenomena. Specifically, we investigate the generation of point defects, which constitutes the atomic-scale mechanism for the observed ionic diffusion. In fluorite-structured ceramics, the "liquid-like" anionic diffusion is enabled by the massive generation and exchange of oxygen Frenkel pairs (OFPs). To investigate this, the evolution of OFPs is quantitatively tracked using the Wigner–Seitz defect analysis method implemented in Ovito software [39]. The stable number of OFPs ( $N_{OFP}$ ) at each temperature is determined by averaging over continuous sampling trajectories (detailed in Section S12 of the Supplementary Materials). As presented in Fig. 13(a), the evolution of $N_{OFP}$ exhibits three distinct stages (LT, MT, and HT), segmented by $T_{Tam}$ and $T_S$ . Notably, an explosive proliferation of OFPs occurs upon exceeding $T_S$ in both RSS and SRO models. This "structural avalanche" corresponds to the pre-melting of the O-sublattice, signifying the transition toward a highly disordered, liquid-like phase—a hallmark of superionic behavior in fluorite materials [16, 49, 78]. Motivated by the Arrhenius analysis performed on O-ions diffusion, we examine the thermodynamic behavior of OFP formation by fitting the temperature dependence of $N_{OFP}$ to an Arrhenius-type relationship as:

$$N_{OFP} = N_0 \exp\left(-\frac{Q_{OFP}}{k_B T}\right) \quad (18)$$

where $A$ is a pre-exponential constant and $Q_{OFP}$ denotes the activation energy for oxygen Frenkel pair formation.

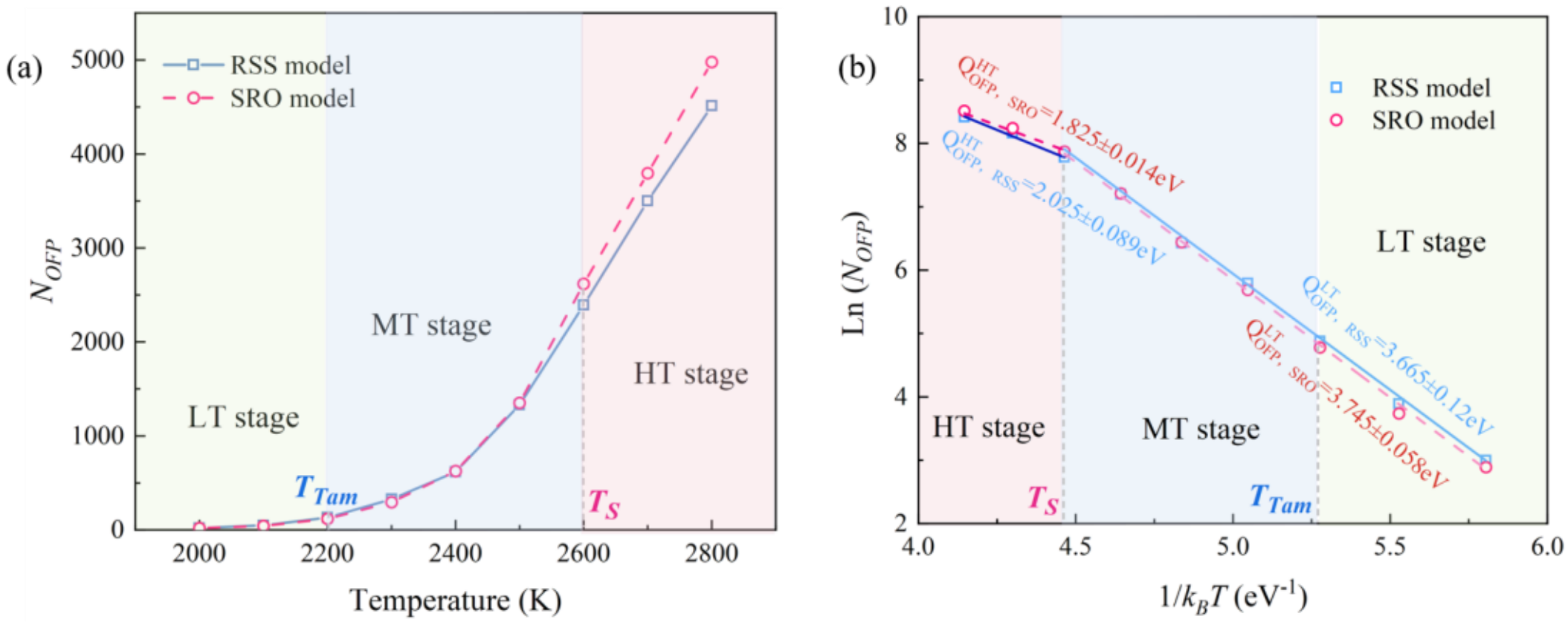

**Fig. 13** (a) Variations of the number of oxygen Frenkel pairs (OFPs) ( $N_{OFP}$ ) as a function of temperature for RSS and SRO HEC models. (b) Arrhenius plots of $\ln(N_{OFP})$ versus $1/k_BT$ . The vertical dashed lines mark the Tammann crossover temperature ($T_{Tam} = 2200$ K) and superionic transition temperature ($T_S = 2600$ K). The plots illustrate that while $N_{OFP}$ is influenced by both $T_{Tam}$ and $T_S$, the thermodynamic activation mechanism of OFP is exclusively determined by $T_S$.

The temperature-dependent variations for the number of OFPs ( $N_{OFP}$ ) and the corresponding Arrhenius-type plots for $\ln(N_{OFP})$ versus $1/k_BT$ are plotted in Fig. (a) and Fig. 13(b), respectively. Clearly, the numbers of OFPs ( $N_{OFP}$ ) increases gradually with rising temperatures, reflecting the progressive accumulation of disorder. Interestingly, unlike the $N_{OFP}$ exhibits two remarkable increments at both $T_{Tam}$ and $T_S$ (Fig. 13(a)), the corresponding activation energy for OFP formation is altered only at $T_S$ (Fig. 13(b)). This suggests that while the structural relaxation initiated at $T_{Tam}$ change the absolute quantity of accumulated defects (as shown in Fig. 9), the fundamental thermodynamic mechanism of defect formation is triggered exclusively by the superionic transition at $T_S$. This finding confirms that the OFP-mediated O-ion diffusion is highly sensitive to the localized caging effects relaxed at $T_{Tam}$, whereas the fundamental thermodynamic shift from localized solid-state defect generation to a pre-melted, disordered mechanism is overwhelmingly governed by the structural superionic threshold at $T_S$. Consequently, $T_S$ serves as the fundamental thermodynamic trigger for defect avalanches, while $T_{Tam}$ acts merely as a kinetic crossover that affects the efficiency of defect migration. Notably, the aforementioned competitive mechanism, characterized by the dynamic competition between the localized SRO-induced blocking and collective superionic mobilization across $T_{Tam}$ and $T_S$, is vividly illustrated in the 'tug-of-war' schematic provided in Section S13 of the Supplementary Materials.

## 5 Conclusions

In this study, the complex competitive mechanisms between the high-entropy effects and superionic transition in $(Th_{0.2},U_{0.2},Pu_{0.2},Np_{0.2},Am_{0.2})O_2$ MA-MOX HEC are systematically investigated. We comprehensively examine the extreme-condition behaviors of HEC after evaluating its fundamental thermodynamic boundaries. These boundaries include the Np-Am SRO determined via hybrid MC/MD optimization, the melting point ($T_m$) calculated via TPS, and superionic transition temperature ($T_S$) identified by tracking the PDF variations of anionic sublattice pre-melting. Drawing on the physical similarities between the superionic anionic sublattice and discorded liquid phases, we pioneer the application of analytical frameworks from amorphous physics to decode this complex structural and kinetic interplay. Specifically, we employ structural mappings including the atomic entropy, volumetric strain, nonaffine squared displacement ( $D^2_{min}$ ), effective viscosity, and structural relaxation time to decode the local microstructure evolutions. These techniques are then integrated with robust statistical metrics, including the spatial inhomogeneity parameter (*h*-factor) and non-Gaussian parameter, to quantitatively analyze the dynamic self-organization of O-ion diffusion. Based on these systemic analyses, the main conclusions are drawn as follows:

(1) SRO-induced blocking and dynamic heterogeneity: The self-clustering of Np and Am acts as the primary SRO structures in the HEC. At lower temperatures, the severe lattice distortions induced by these SRO structures create localized electrostatic and strain-coupled potential wells, exerting a blocking effect that obstructs mobility and induces sluggish diffusion of O-ions. As the system enters the superionic phase, the *h*-factor quantitatively reveals that the SRO model strongly resists thermal homogenization, maintaining a significantly higher spatial heterogeneity of mobile O-ions compared to RSS model. This structural resistance forces O ions to bypass the SRO regions, driving a dynamic self-organization that establishes preferential, string-like active diffusion sites and ultimately leads to confinement-enhanced mobility.

(2) Topological evolution and kinetic inversion: The superionic transition triggers

profound structural and thermodynamic anomalies, characterized by the anionic sublattice pre-melting and massive oxygen Frenkel pair (OFP) generation. Topologically, this transition is marked by the rapid collapse of intermediate-temperature Voronoi polyhedra and the absolute dominance of <0,4,4,0>-indexed polyhedra. These <0,4,4,0>-indexed polyhedra spontaneously self-organize into highly interconnected, string-like networks. Acting as the structural carriers for collective liquid-like motion for O-ions, these topological networks drive the anomalous kinetic inversion observed in structural relaxation times.

(3) Three-stage thermal evolution of O-ion transport: The thermal evolution of the anionic sublattice is partitioned into three distinct kinetic and thermodynamic stages by the Tammann crossover temperature ($T_{Tam} = 2200\,\mathrm{K}$) and superionic transition temperature ($T_S = 2600$ K). In the low-temperature stage ($T < T_{Tam}$), rigid SRO structures impose severe local blocking, restricting O-ion mobility to localized solid-state dynamics. In the mid-temperature stage ($T_{Tam} < T < T_S$), thermal activation alleviates these structural constraints, inducing a transitional state dominated by intense dynamical heterogeneity. Ultimately, in the high-temperature stage ($T > T_S$), the system undergoes a structural avalanche where the thermodynamic driving force of the superionic transition completely overrides the localized SRO confinement, serving as the fundamental trigger for massive defect generation and liquid-like diffusion.

**CRediT authorship contribution statement**

**F.N. Zhuang:** Investigation, Conceptualization, Validation, Writing – original draft, Writing – review & editing. **G.S. Yan:** Writing – review & editing, Supervision. **H. Chen**: Supervision & Funding acquisition. **Y. Zhang**: editing & review. **W.S. Yu**: Writing – original draft, Writing – review & editing, Supervision, Funding acquisition. **M.L. Xu**: Supervision & editing. **S.P. Shen**: Writing – review & editing, Supervision, Funding acquisition.

**Declaration of Competing Interest**

The authors declare that they have no known competing financial interests or personal relationships that could have appeared to influence the work reported in this paper.

**Acknowledgments**

The authors would like to express the gratitude for the support of NSFC (Grant No: 12202341 and 12527801). This study is supported by project of the Ministry of Industry and Information Technology.